\documentclass[a4paper,australian,aps,reprint,twocolumn,twoside,preprintnumbers,showkeys,showpacs,final]{revtex4-1}
\usepackage{amsmath}
\usepackage{amssymb}
\usepackage{siunitx}
\usepackage{color}
\usepackage{natbib}
\usepackage{graphicx}
\usepackage{placeins}
\usepackage{hyperref} 
\usepackage{ulem}

\usepackage{float}
\graphicspath{ {./} }

\DeclareMathOperator{\Tr}{Tr}

\begin{document}

\preprint{ADP-18-16/T1064}

\title{Gluon propagator on a centre-vortex background}
\author{James C. Biddle}
\author{Waseem Kamleh}
\author{Derek B. Leinweber}
\affiliation{Centre for the Subatomic Structure of Matter, Department of Physics, The University of Adelaide, SA 5005, Australia}

\begin{abstract}
The impact of $SU(3)$ centre vortices on the gluon propagator in Landau gauge is investigated on original, vortex-removed and vortex-only lattice gauge field configurations. Vortex identification is found to partition the gluon propagator into short range strength on the vortex removed configurations and long range strength on the vortex only configurations. The effect of smoothing vortex-only configurations is also studied, and a regime for recovering the form of the smoothed original propagator from vortex-only configurations is introduced. The results reinforce the significance of centre vortices in a fundamental understanding of QCD vacuum structure.
\end{abstract}

\pacs{12.38.Gc,12.38.Aw,14.70.Dj}
\keywords{Gluon propagator; centre vortices; Lattice QCD; Landau gauge}

\maketitle
\section{Introduction}
There is now significant evidence supporting the role of centre vortices in confinement and dynamical chiral symmetry breaking in QCD. It is well established that for $SU(2)$ gauge theories, centre vortices account for both of these properties~\cite{Bowman:2008qd,DelDebbio:1996lih,Greensite:2003bk,DelDebbio:1998luz} and thus it would seem intuitive that the same would hold for SU(3) theories as well. Previous work has successfully shown that vortex removal in $SU(3)$ corresponds to a loss of dynamical mass generation and string tension~\cite{OMalley:2011aa,Trewartha:2015nna}. However, complete recovery of these properties on vortex only backgrounds has proven difficult~\cite{Langfeld:2003ev}.

In this paper we explore the behaviour of the Landau gauge gluon propagator~\cite{Zwanziger:1991gz,Cucchieri:1999sz,Leinweber:1998uu} on original, vortex-removed and vortex-only configurations. Suppression of the gluon propagator after vortex removal has been previously demonstrated~\cite{Langfeld:2001cz,Bowman:2010zr}. However, the scalar propagator has not yet been examined on a vortex only background.

Studies of the overlap fermion quark propagator  have found that by starting from the vortex-only fields, which consist only of the centre elements of $SU(3),$ and through the application of a link smoothing algorithm, one is able to reproduce all the salient features of QCD, including confinement~\cite{Trewartha:2015ida}, dynamical mass generation~\cite{Trewartha:2015nna}, and the low-lying hadron spectrum~\cite{Trewartha:2017ive}. In this work, we extend this line of investigation by examining the effect of cooling and smearing on the gluon propagator obtained from vortex-modified gauge field configurations. The purpose of this investigation is to identify whether is is possible to reproduce a gluon propagator indicative of confining behaviour on vortex-only configurations.
\vspace{-4pt}
\section{Landau Gauge Gluon Propagator}\label{sec:GluonProp}
\vspace{-2pt}
The momentum space gluon propagator on a finite lattice with four-dimensional volume $V$ is given by
\begin{equation}
D_{\mu\nu}^{ab}(p) \equiv \frac{1}{V}\left \langle A^a_\mu (p)\,A^b_\nu(-p)\right\rangle \, , \label{eq:gluonProp}
\end{equation}
where $A^a_\mu$ are the Hermitian gluon fields (see the Appendix for more details). In the continuum, the Landau-gauge momentum-space gluon propagator has the following form~\cite{Leinweber:1998im,Bonnet:2001uh}
\begin{equation}
D^{ab}_{\mu\nu}(q) = \left ( \delta_{\mu\nu} - \frac{q_\mu q_\nu}{q^2} \right )\,\delta^{ab}\,D(q^2) \, ,
\end{equation}
where $D(q^2)$ is the scalar gluon propagator.  Contracting Gell-Mann index $b$ with $a$ and
Lorentz index $\nu$ with $\mu$ one has
\begin{equation}
D^{aa}_{\mu\mu}(q) = (4-1)\,(n_c^2-1)\,D(q^2) \, ,
\end{equation}
such that the scalar function can be obtained from the gluon propagator via
\begin{equation}
D(q^2) = \frac{1}{3(n_c^2-1)}\,D^{aa}_{\mu\mu}(q) \, ,
\label{eq:scalarProp}
\end{equation}
where $n_c = 3$ is the number of colours.

As the lattice gauge links $U_\mu(x)$ naturally reside in the fundamental representation of $SU(3),$ it is convenient to work with the corresponding $3\times 3$ matrix representation of the gauge potential $A_\mu  = A^a_\mu\,(\lambda_a/2),$ where $\lambda_a$ are the eight Gell-Mann matrices. Using the orthogonality relation $\Tr(\lambda_a\lambda_b) = \delta_{ab}$ for the Gell-Mann matrices, it is straightforward to see that
\begin{equation}
2\Tr(A_\mu\,A_\mu) = A^a_\mu A^a_\mu,
\end{equation}
which can be substituted into equation~\ref{eq:scalarProp} to obtain the final expression for the lattice scalar gluon propagator,
\begin{equation}
D(p^2) = \frac{2}{3\,(n_c^2-1)\,V}\big\langle {\rm Tr}\, A_\mu(p)\,A_\mu(-p) \big\rangle \,. \label{eq:scalarProp2}
\end{equation}

Following the formalism of Ref.~\cite{Leinweber:1998im}, we calculate
the lattice gluon propagator using the mid-point
definition of the gauge potential in terms of the lattice link
variables~\cite{Alles:1996ka},
\begin{multline}
A_\mu(x+\hat{\mu}/2)=\frac{1}{2i}\, \left (U_\mu(x)-U^{\dagger}_\mu(x)\right ) \\
-\frac{1}{6 i}\, {\rm Tr}\left (U_\mu(x)-U^{\dagger}_\mu(x) \right ) + {\cal O}(a^2)\,.  \label{eq:midPoint}
\end{multline}
In this definition of the gauge potential we have absorbed the lattice spacing $a$ and the strong coupling constant $g$ into $A_\mu$, such that $ga\,A_\mu\rightarrow A_\mu$. The gluon fields $U_\mu(x)$ are first gauge-fixed by maximizing an $\mathcal{O}(a^{2})$-improved functional using a Fourier-accelerated algorithm~\cite{Davies:1987vs,Bonnet:1999mj,Roberts:2010cz}. The gauge potential in momentum space is then obtained by taking the discrete Fourier transform, 
\begin{equation}
A_\mu(p) = \sum_x e^{-ip\cdot(x+\hat{\mu}/2)}\, A_\mu(x+\hat{\mu}/2).
\end{equation}

The gauge fields used in this analysis are created using the $\mathcal{O}(a^2)$-improved L\"uscher and Weisz action~\cite{Luscher:1984xn}. It is known that the continuum propagator has the form 
\begin{equation}
D(p^2)=\frac{1}{p^2}\, ,
\end{equation}
as $p^2\rightarrow\infty$. To preserve this behaviour on the lattice, it is necessary to make use of the momentum variable $q_\mu$ defined by the tree-level form of the $\mathcal{O}(a^2)$-improved gluon propagator~\cite{Bonnet:2001uh,Bowman:2004jm}.
\begin{equation}
q_\mu = \frac{2}{a}\sqrt{\sin^2\left(\frac{p_\mu a}{2}\right)+\frac{1}{3}\sin^4\left(\frac{p_\mu a}{2}\right)}\, ,
\end{equation}
where $p_\mu$ are the usual lattice momentum variables
\begin{equation}
p_\mu = \frac{2\pi\,n_\mu}{aN_\mu}, ~~n_\mu \in \left(-\frac{N_\mu}{2},\frac{N_\mu}{2}\right]\, .
\end{equation}
This choice of a $\mathcal{O}(a^2)$-improved action, gauge-fixing functional~\cite{Bonnet:1999mj} and momentum ensures that we reduce the sensitivity of the gluon propagator to finite lattice-spacing effects~\cite{Bonnet:2001uh}.

We follow the tradition of examining $q^2D(q^2)$ such that at large $q^2$ we observe $q^2D(q^2)$ trending towards a constant. We then renormalise such that $q^2D(q^2)=1$ for $qa = 3.0$ on the original configurations, and apply this same renormalisation factor to all subsequent vortex-modified propagators.


\section{Centre Vortex Projection}

\begin{figure}[tb]
\centering
\includegraphics[width=\linewidth]{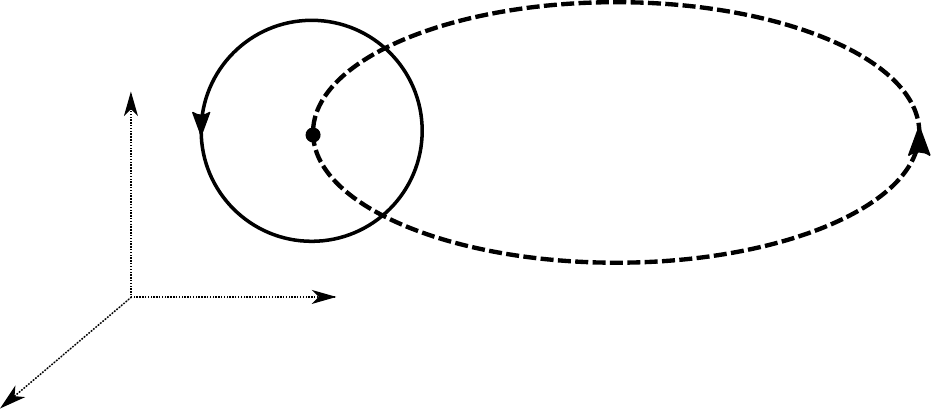}
\caption{\label{fig:vortex}An illustration of a Wilson loop in the vertical plane (solid line) that is pierced by a thin centre vortex (dashed line) at a single point. Only three of the four Euclidean dimensions are shown, such that the intersection with the two-dimensional vortex surface appears as a line.}
\end{figure}


In the centre vortex model of confinement~\cite{'tHooft:1977hy,'tHooft:1979uj}, centre vortices are associated with regions of an $SU(N)$ gauge field that have a non-trivial topology. In a four-dimensional Euclidean space-time, the physical centre vortices present in the QCD vacuum form three-dimensional volumes. These physical or thick vortices are distinguished from the concept of thin vortices. In the continuum (in four dimensions), a thin vortex forms a two-dimensional surface. The thin vortex surface can be related to the region occupied by a corresponding thick vortex, where the former is characterised by an infinitesimal profile while the latter has a finite extent~\cite{Engelhardt:1999xw,Bertle:2000ap}.

A key property of the thin vortices is that Wilson loops which enclose a vortex line acquire a non-trivial centre phase $e^{n2\pi i/N},\,n=0,1,\ldots,N\!-\!1.$  An illustration of a Wilson loop that is pierced by a thin centre vortex is shown in Figure~\ref{fig:vortex}. Three of the four space-time dimensions are shown. The thin vortex is represented by the dashed line. In this instance, the extent of the two-dimensional vortex surface in the fourth dimension is suppressed, and only the intersection with the three-dimensional space is shown. The oriented solid circle represents a Wilson loop which lies within the vertical plane, and is pierced by the vortex line at a single point. The vortex intersection causes the Wilson loop to acquire a non-trivial centre phase.

On the lattice, the trace of the elemental plaquettes represent the smallest non-trivial Wilson loops. We decompose the $SU(3)$ lattice gauge links 
\begin{equation}
U_{\mu}(x) = Z_{\mu}(x)\cdot R_{\mu}(x),
\end{equation}
in such a way that all vortex information is captured in the field of centre-projected elements $Z_{\mu}(x)$,
\begin{equation}
Z_\mu(x) = e^{k2\pi i/3}\,I,\quad k\in \{ -1,0,+1 \},
\end{equation}
with the remaining short-range fluctuations described by the vortex-removed field $R_{\mu}(x).$ The centre-projected plaquettes in the $Z_{\mu}(x)$ configurations with a nontrivial flux around the boundary form the thin vortices (or P-vortices) that are embedded within the thick vortices of the original Monte Carlo configurations. Note that while a fundamental plaquette is sufficient to detect a thin vortex on the lattice, a thick vortex requires a Wilson loop of finite size to acquire the centre element~\cite{DelDebbio:1998luz}.

To identify centre vortices in the Monte Carlo generated configurations we follow the maximal centre gauge (MCG) centre projection procedure as described in Refs.~\cite{Montero:1999by,Faber:1999sq}. We seek a gauge transformation $\Omega(x)$ that minimises
\begin{equation}
||U_{\mu}^\Omega(x)-Z_\mu(x)||\, ,
\end{equation}
where $Z_\mu(x)$ are the centre elements of $SU(3).$ This transformation is performed by maximising the so-called ``mesonic" functional\, \cite{Langfeld:2003ev}
\begin{equation}
\label{eq:MCGfunctional}
R=\frac{1}{V\,N_{\text{dim}}\,n_c^2}\sum_{x,\mu}|\text{Tr}\,U_{\mu}^\Omega(x)|^2\, .
\end{equation}
Once the configurations are fixed to maximal centre gauge, each link can be projected onto the nearest centre element. We define these projected configurations $Z_\mu(x)$ as the vortex-only configurations. This projection also allows us to define the vortex removed configurations
\begin{equation}
R_\mu(x) = Z_\mu^\dagger(x)\,U_\mu(x)\, .
\end{equation}
Hence, we refer to the three different gauge field ensembles created as follows:
\begin{enumerate}
\item Original `untouched' fields, $U_{\mu}(x),$

\item Projected vortex-only fields, $Z_{\mu}(x),$

\item Vortex-removed fields, $R_\mu(x) = Z^{\dagger}_{\mu}(x)\,U_{\mu}(x).$
\end{enumerate}


\section{Results}
\subsection{Survey of configurations}
\begin{figure}[tb]
\centering
\includegraphics[width=\linewidth]{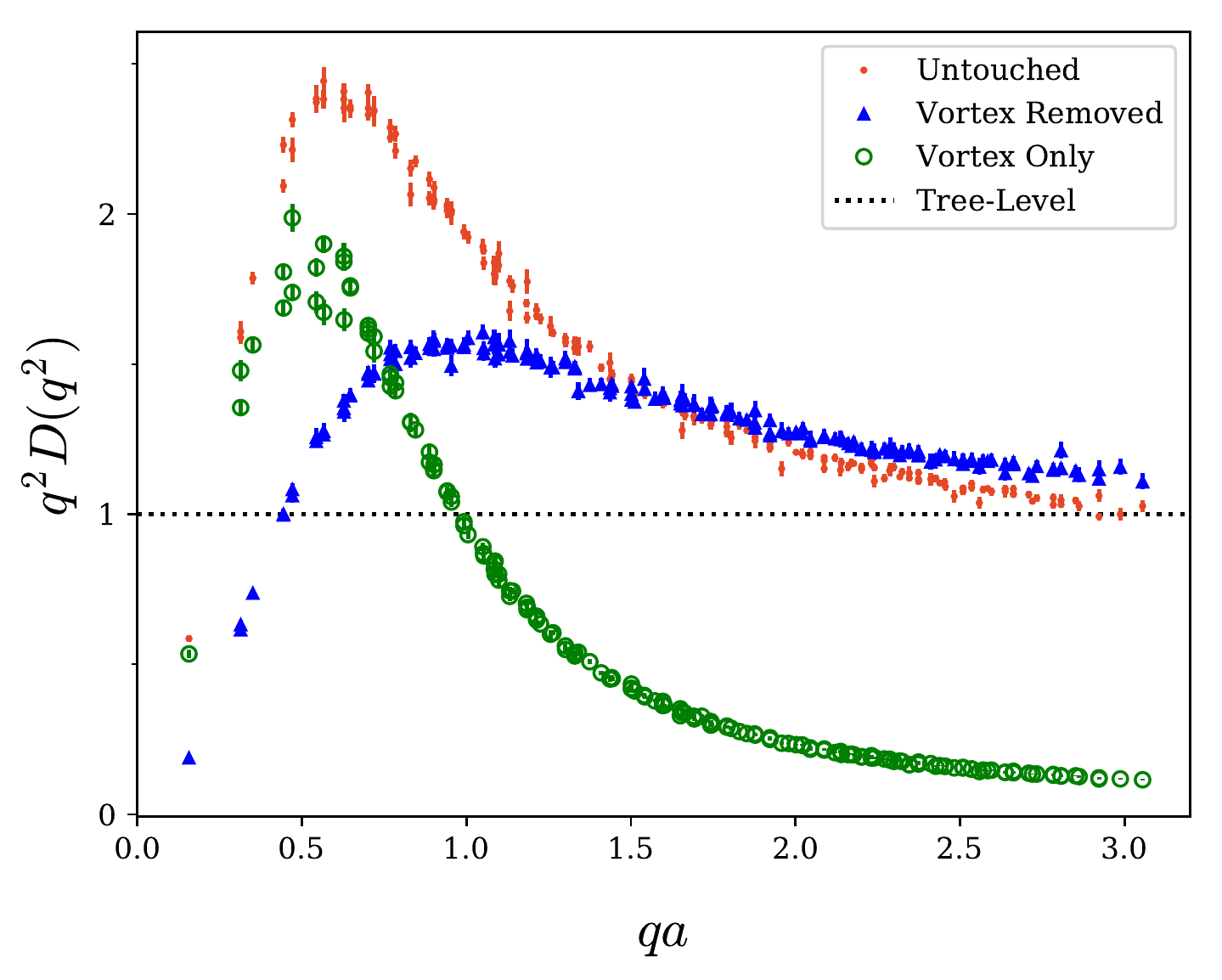}
\caption{\label{nocoolsum}The gluon propagator calculated from the original untouched (red dots), shown with the vortex removed (blue triangles) and vortex only (green open circles) results. Here, the renormalisation factor for the vortex removed and vortex only propagators is chosen to be the same as for the untouched propagator.}
\end{figure}
\begin{figure}[tb]
\centering
\includegraphics[width=\linewidth]{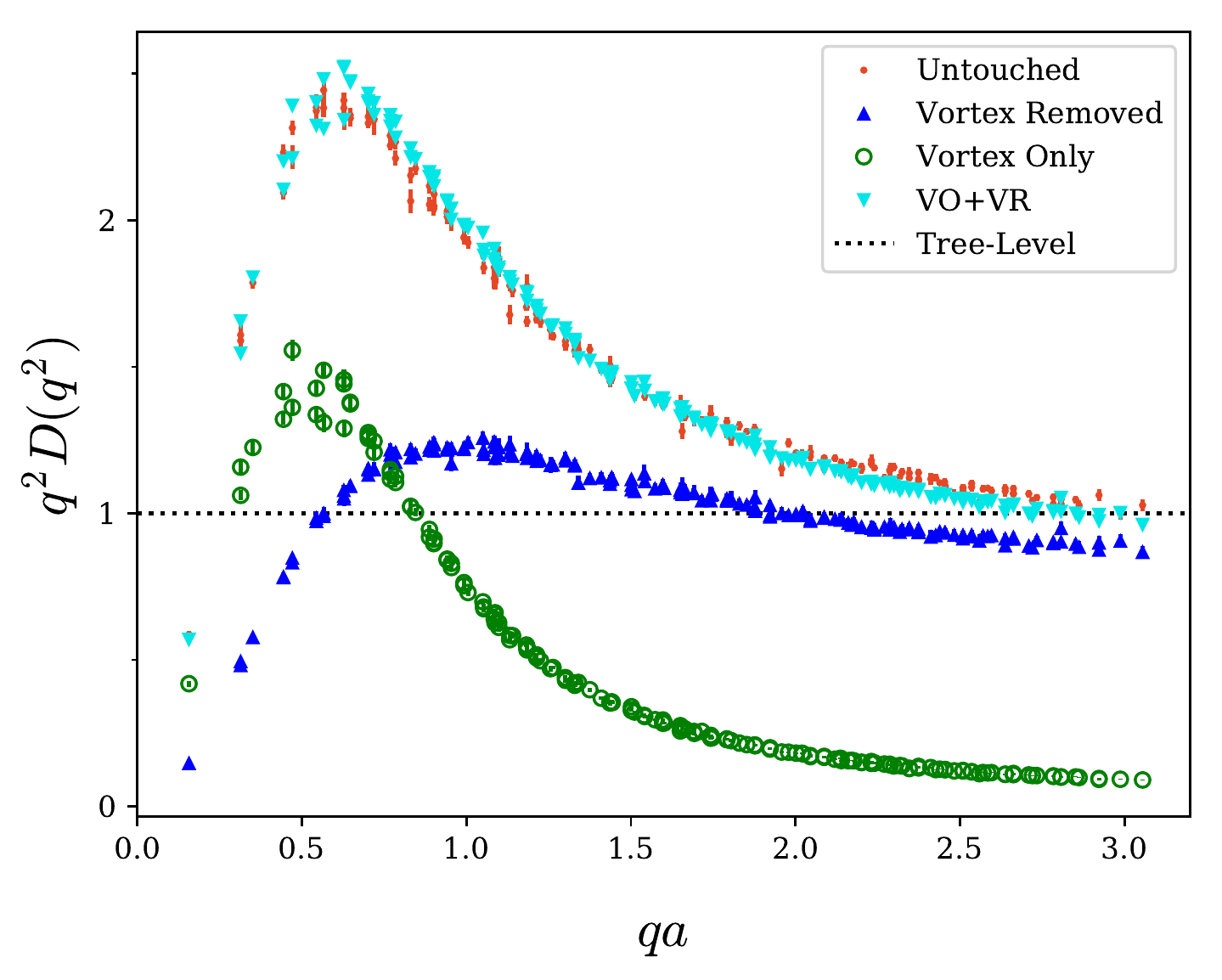}
\caption{\label{nocoolsum2}The gluon propagator from the original untouched ensemble as in Fig.~\ref{nocoolsum}, now shown with the independently renormalised sum (cyan triangles) of the vortex removed and vortex only propagators. The two vortex modified propagators are also shown, but here their renormalisation factor is chosen to be the same as for the summed propagator.}
\end{figure}
We calculate the gluon propagator on 100 configurations of a $20^3\times 40$ $SU(3)$ lattice with spacing $a=0.125\,\si{fm}$, as used in Refs.~\cite{Trewartha:2015nna,OMalley:2011aa}. Following the procedure of Ref.~\cite{Bonnet:2001uh,Leinweber:1998im} all results are plotted after a momentum half-cut and a cylinder cut of radius $pa=2$ lattice units have been performed. Additionally, we can take advantage of the rotational symmetry of the scalar propagator to perform $Z(3)$ averaging over the Cartesian coordinates. This means that we average over all points with the same Cartesian radius; for example, we would average across the points $(n_x,n_y,n_z)=(2,1,1),\,(1,2,1)$ and $(1,1,2)$. 

Calculating the scalar propagator on untouched, vortex-removed and vortex only configurations gives the results illustrated in Fig.~\ref{nocoolsum}. As discussed at the end of section~\ref{sec:GluonProp}, we renormalise such that $q^2D(q^2)=1$ for $qa = 3.0$ on the original configurations, and apply this same renormalisation factor to the vortex removed and vortex only propagators. The vortex removed configurations display the expected behaviour, with vortex removal corresponding to significant infrared suppression of the propagator when compared to the untouched propagator, in agreement with the results of Ref.~\cite{Bowman:2010zr}. The increased roughness of the gauge fields after vortex removal is evidenced by the enhancement of the propagator at large $q$. This reflects the increase in short-distance fluctuations that have been introduced to the gauge fields by the vortex removal procedure.

It is interesting to note that the vortex only propagator retains approximately two thirds of the untouched propagator's peak strength. This is comparable to previous work showing partial recovery of the string tension on vortex only configurations~\cite{Trewartha:2015ida,Trewartha:2017ive,Langfeld:2003ev,Stack:2002sy}. Despite only recovering a portion of the original strength, the infrared peak is still considerably greater than the peak observed in the vortex removed propagator. The loss of strength is most likely in part because of the known imperfections in the vortex identification algorithm that results in some vortex matter remaining in the vortex removed configurations. The vortex only configurations also exhibit a loss of short range strength, due to the absence of the high frequency modes that are instead contained within the vortex removed field.

If we sum the vortex only and vortex removed propagators and independently renormalise such that $q^2\,D(q^2)=1$ at $qa=3.0$, we obtain the result shown in Fig.~\ref{nocoolsum2}. Here we observe agreement between the untouched and summed propagators. This indicates that vortex modification effectively partitions the lattice configuration into short range physics on the vortex removed configurations and long range physics on the vortex only configurations, up to errors in the vortex identification procedure.

This partitioning is expected if the vortex removed and vortex only configurations are orthogonal. To see how this behaviour emerges, suppose that we can decompose the gluon field $A_\mu$ into two independent fields as follows
\begin{equation}
  A_\mu(p) = B_\mu(p) + C_\mu(p).
\end{equation}
In the context of this work, we associate $B_\mu$ with the background field of short-range gluon fluctuations and $C_\mu$ with the centre vortex field.
Note also that if $B$ and $C$ are in Landau gauge then so is $A.$ Using this partitioning it follows that the gluon propagator for $A$ can be written as the sum of the respective gluon propagators for $B$ and $C,$
\begin{align}
  D^{A}_{\mu\nu}(p) &= \frac{1}{V}\langle A_\mu(p) \, A_\nu(-p) \rangle\nonumber \\
  &= \frac{1}{V}\Big(\langle B_\mu(p)B_\nu(-p)\rangle + \langle C_\mu(p)C_\nu(-p)\rangle\nonumber\\
   &~~~~~~~+ \langle B_\mu(p)C_\nu(-p)+ C_\mu(p)B_\nu(-p) \rangle\Big)\nonumber \\
  &= D^{B}_{\mu\nu}(p) + D^{C}_{\mu\nu}(p),\label{eq:partition}
\end{align}
where we have made use of the fact that $B$ and $C$ represent orthogonal degrees of freedom in the gauge field and hence in the ensemble average the cross-correlations should vanish.

To elucidate the connection to the unitary formulation of the lattice gauge links, we suppose that we can transform $A$ to an ``ideal centre gauge'' such that in lattice units the field $C$ consists purely of centre phases,
\begin{equation}
C_\mu(x) = k\,\frac{2\pi}{3} I,\quad k\in \{-1,0,+1\}.
\end{equation}
On a continuous manifold we can write the Wilson line corresponding to a lattice link as a path-ordered exponential,
\begin{equation}
U_{\mu}(x) = \mathcal{P} e^{i\int_0^1 d\lambda \,A_{\mu}(x+\lambda\hat{\mu})}.
\end{equation}
The lattice midpoint approximation replaces the integral as follows,
\begin{equation}
U_{\mu}(x) = e^{iA_{\mu}(x+\hat{\mu}/2)}.
\end{equation}
As $A = B + C$ it immediately follows that we can write
\begin{equation}
U_{\mu}(x) = e^{i B_{\mu}(x+\hat{\mu}/2)} \, e^{i C_{\mu}(x+\hat{\mu}/2)},
\end{equation}
noting that in our ideal centre gauge $[B,C] = 0$ so the Baker-Campbell-Haussdorff relation is trivial. Identifying
\begin{equation}
Z_{\mu}(x) = e^{i C_{\mu}(x+\hat{\mu}/2)}
\end{equation}
as the vortex-projected field, and
\begin{equation}
R_{\mu}(x) = e^{i B_{\mu}(x+\hat{\mu}/2)}
\end{equation}
as the background remainder field we thus recover the decomposition of the links used herein,
\begin{equation}
U_{\mu}(x) = Z_{\mu}(x)\cdot R_{\mu}(x).
\end{equation}
In practise, on the lattice the maximal centre gauge fixing that is implemented will differ from the ideal centre gauge postulated here due to apparent numerical difficulties in simultaneously identifying all vortex matter within an $SU(3)$ gauge field. What this means is that the projected field $Z$ may not capture all of the vortex matter such that there is some non-trivial topological structures that remain in the background field $R.$ The infrared enhancement in the vortex removed results in Fig.~\ref{nocoolsum} suggests this is the case.

\subsection{Smoothing}\label{sec:Smoothing}

It has previously been shown that smoothing is necessary to obtain agreement between the untouched and vortex only string tension, mass function and instanton content~\cite{Trewartha:2015ida,Trewartha:2015nna,Trewartha:2017ive}. Motivated by these results, we now investigate the effect of both $\mathcal{O}(a^4)$-improved cooling~\cite{BilsonThompson:2003zi} and over-improved stoutlink smearing~\cite{Moran:2008ra}. Following the results of Ref.~\cite{Moran:2008ra}, the over-improved smearing parameters are $\rho=0.06$ and $\epsilon=0.25$ to best preserve the size of instantons on the lattice. To accomplish a similar preservation of topological objects under cooling, we used the three-loop improved algorithm as described in Ref.~\cite{BilsonThompson:2003zi}.

\begin{figure}[tb]
\centering
\includegraphics[width=\linewidth]{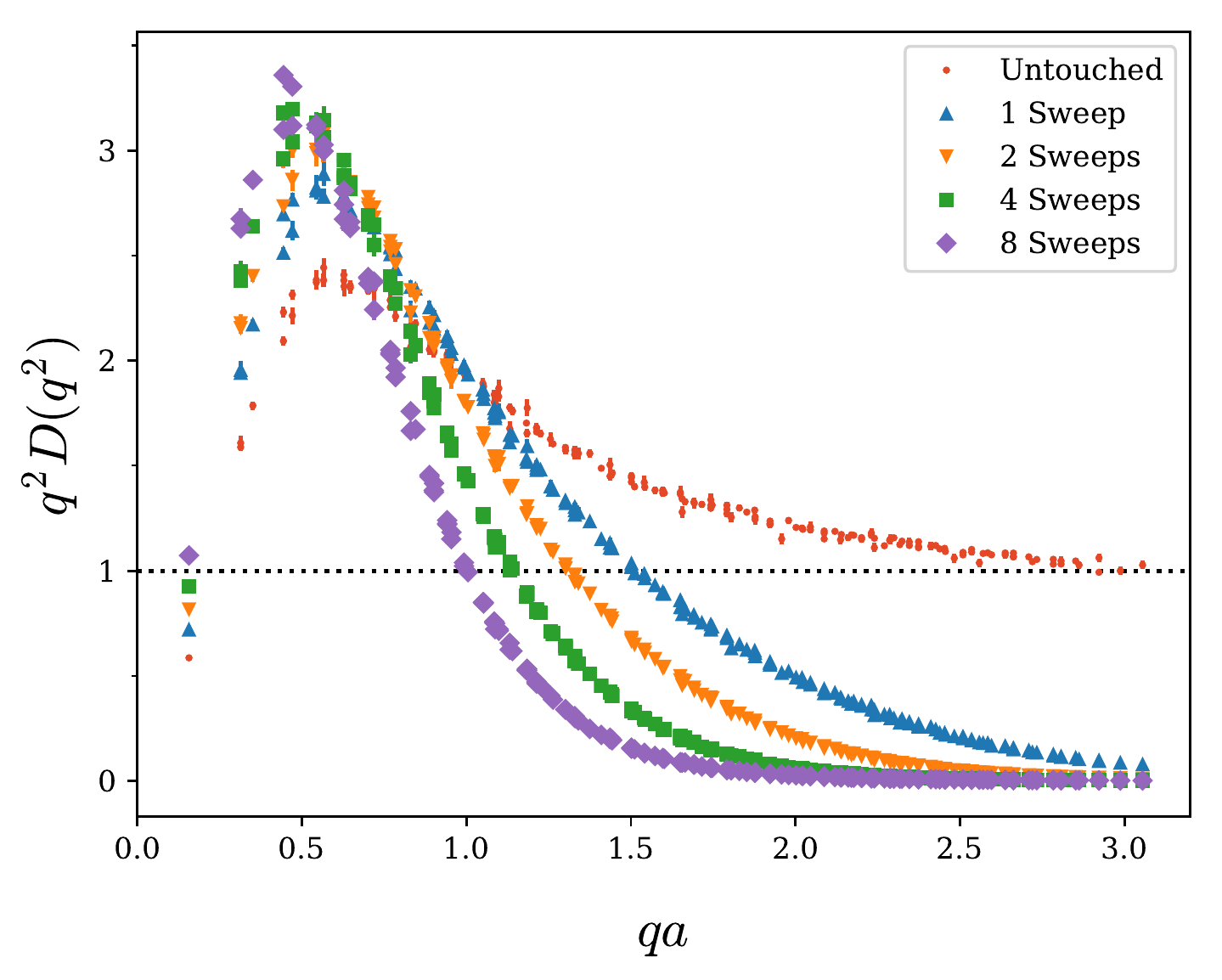}
\caption{\label{1to10sweepscooling}Comparison of the gluon propagator on the untouched configurations after cooling. For clarity we have selected a sample of sweeps between 1 and 8.}
\end{figure}
We first plot the untouched propagator after 0, 1, 2, 4 and 8 sweeps of cooling in Fig.~\ref{1to10sweepscooling}. In gauge fixing, each sweep has been preconditioned by the Landau gauge transformation of the prior sweep in descending order (i.e. the transformation for sweep 10 preconditions sweep 9). This preconditioning is done to ensure that the Landau gauge functional is near the same local minima for each cooling sweep. We observe the expected removal of short distance fluctuations that is typical of smoothing, resulting in a suppressed propagator at large $q$. This is complemented by an amplification in the infra-red region which can be attributed to the increase in low momentum modes arising from the smoothing of the gauge fields.

\begin{figure}[tb]
\centering
\includegraphics[width=\linewidth]{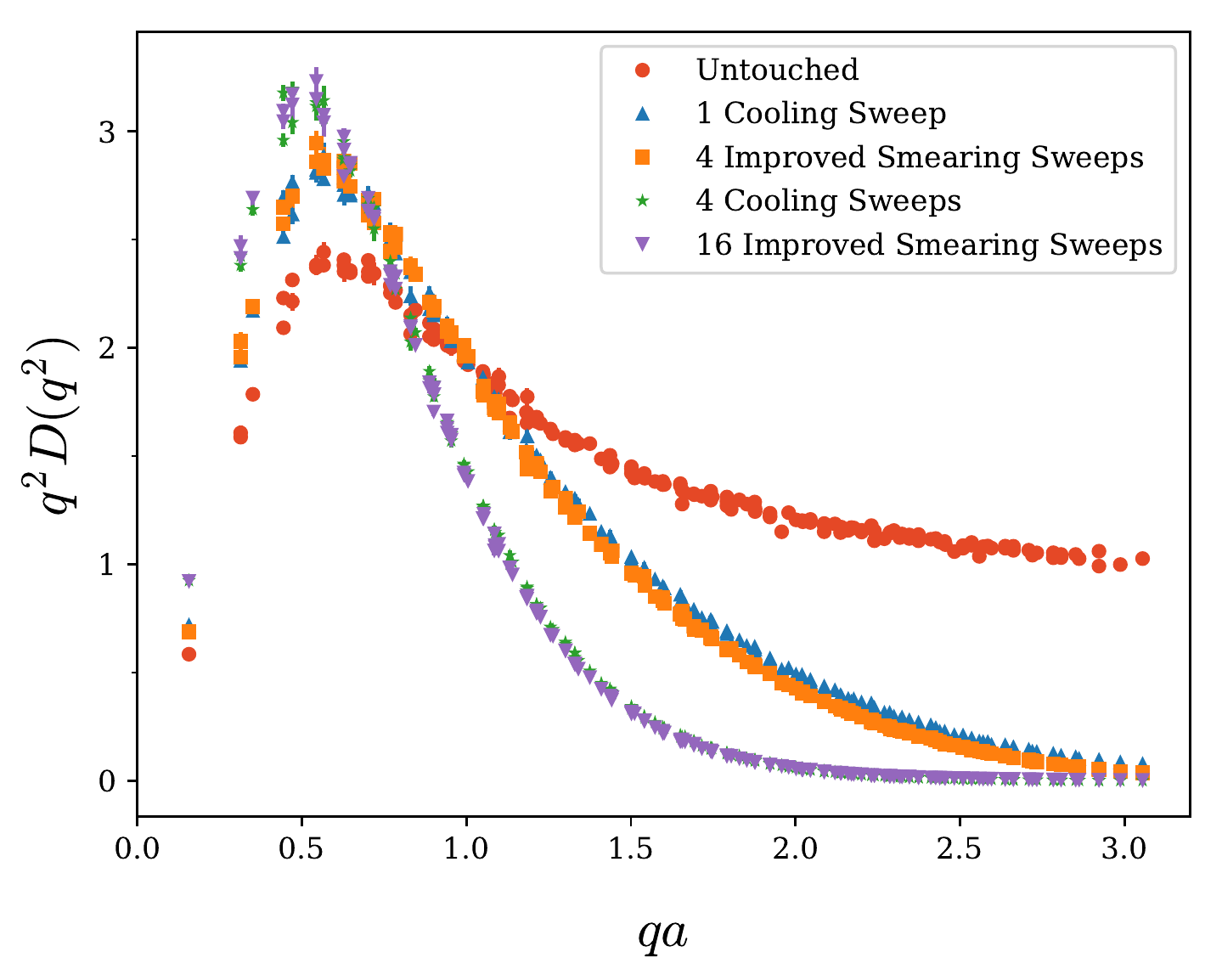}
\caption{\label{smearcoolcomp}The gluon propagator after cooling or improved smearing. We see that the shape of the plot changes minimally between the smoothing routines. However cooling requires fewer sweeps to produce the same effect when compared to smearing.}
\end{figure}
To compare the effects of cooling and over-improved smearing, the untouched gluon propagator is plotted in Fig.~\ref{smearcoolcomp} after either over-improved smearing or cooling. By comparing the smeared and cooled propagator we can see that cooling has a more rapid effect, related to the well-known fast removal of action from the lattice. The qualitative shape of the propagator remains the same however, and it can be seen that, for example, 4 smearing sweeps produces a propagator remarkably similar to 1 cooling sweep. More generally, we observe that in regards to the shape of the propagator, $n_{\text{sm}}\approx4\,n_{\text{cool}}$. Following the observation made in Ref.~\cite{Thomas:2014tda} that the number of over-improved stoutlink smearing sweeps is related to the gradient flow time by
\begin{equation}
t\approx\rho\,n_{\text{sm}}\, ,
\end{equation}
we deduce that the relationship between gradient flow time and cooling is
\begin{equation}
t\approx0.24\,n_{\text{cool}}\,.
\end{equation}
It is well understood that smoothing alters the vortex background, and based on previous work~\cite{Cais:2008za,Trewartha:2015ida,DelDebbio:1998luz} we anticipate that the vortices identified on smoothed configurations would differ to those identified on the unsmoothed configurations. We therefore perform vortex identification only on the original configurations, with smoothing then being performed independently on the untouched, vortex-only and vortex-removed configurations.
We choose to use cooling as the smoothing algorithm for the results presented in this paper, however it is worth noting that similar results can be obtained with the use of over-improved smearing. 
\subsection{Role of centre vortices}

\begin{figure}[tb]
\centering
\includegraphics[width=\linewidth]{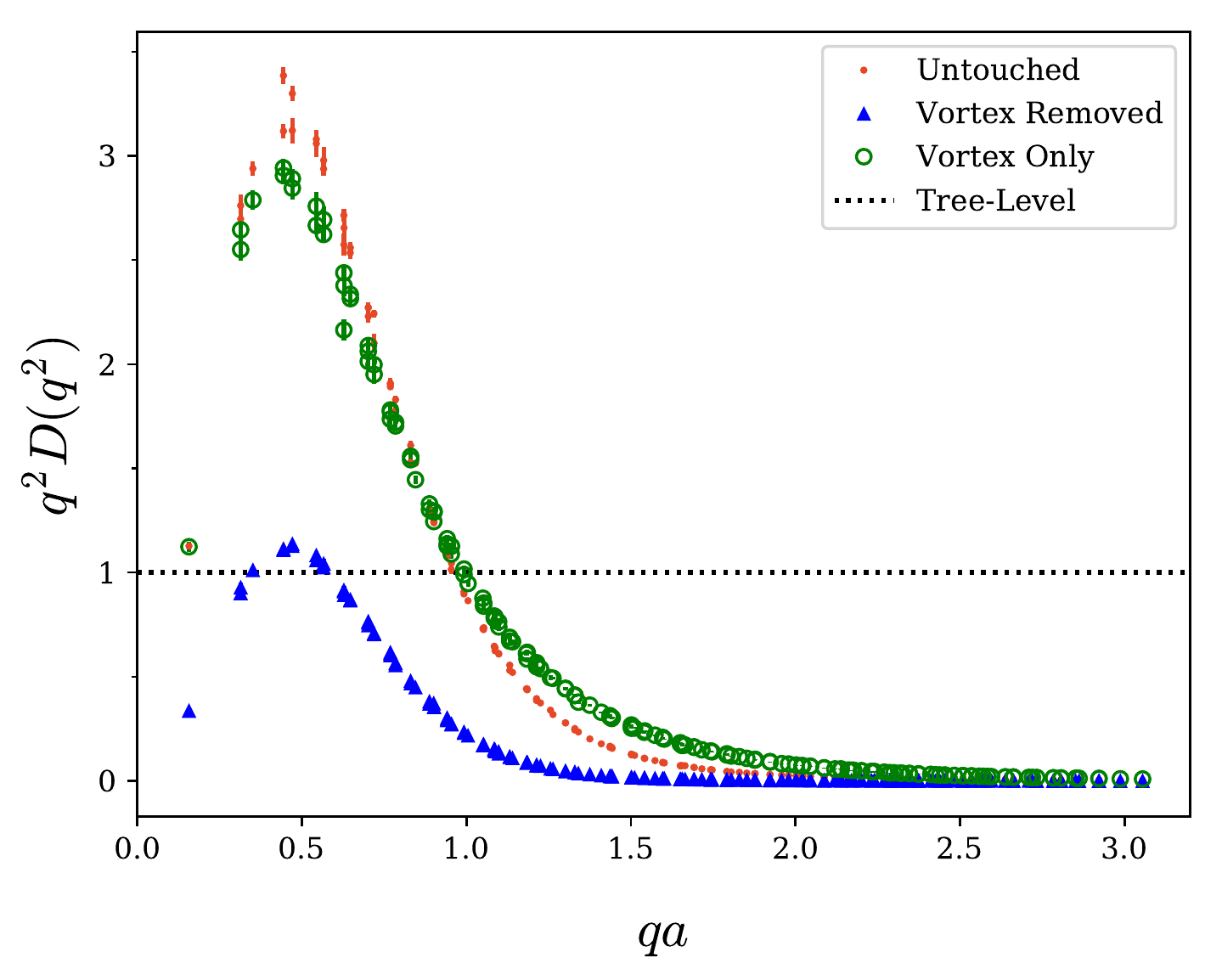}
\caption{\label{10sweepscooling}The gluon propagator calculated on the three ensembles after 10 sweeps of cooling. We now observe an improved agreement between the untouched and vortex only propagators.}
\end{figure}  
After performing 10 sweeps of cooling on the untouched, vortex-removed and vortex-only ensembles, we obtain the results shown in Fig.\,\ref{10sweepscooling}. As is typical of cooling, the removal of short range structures means that all three ensembles tend to zero as $q\rightarrow\infty$. There is now a noticeable improvement in the agreement between the untouched and vortex only configurations; however there is still a difference present, especially in the $qa\approx0.5$ and $qa\approx1.5$ regions.

\begin{figure}[tb]
\centering
\includegraphics[width=\linewidth]{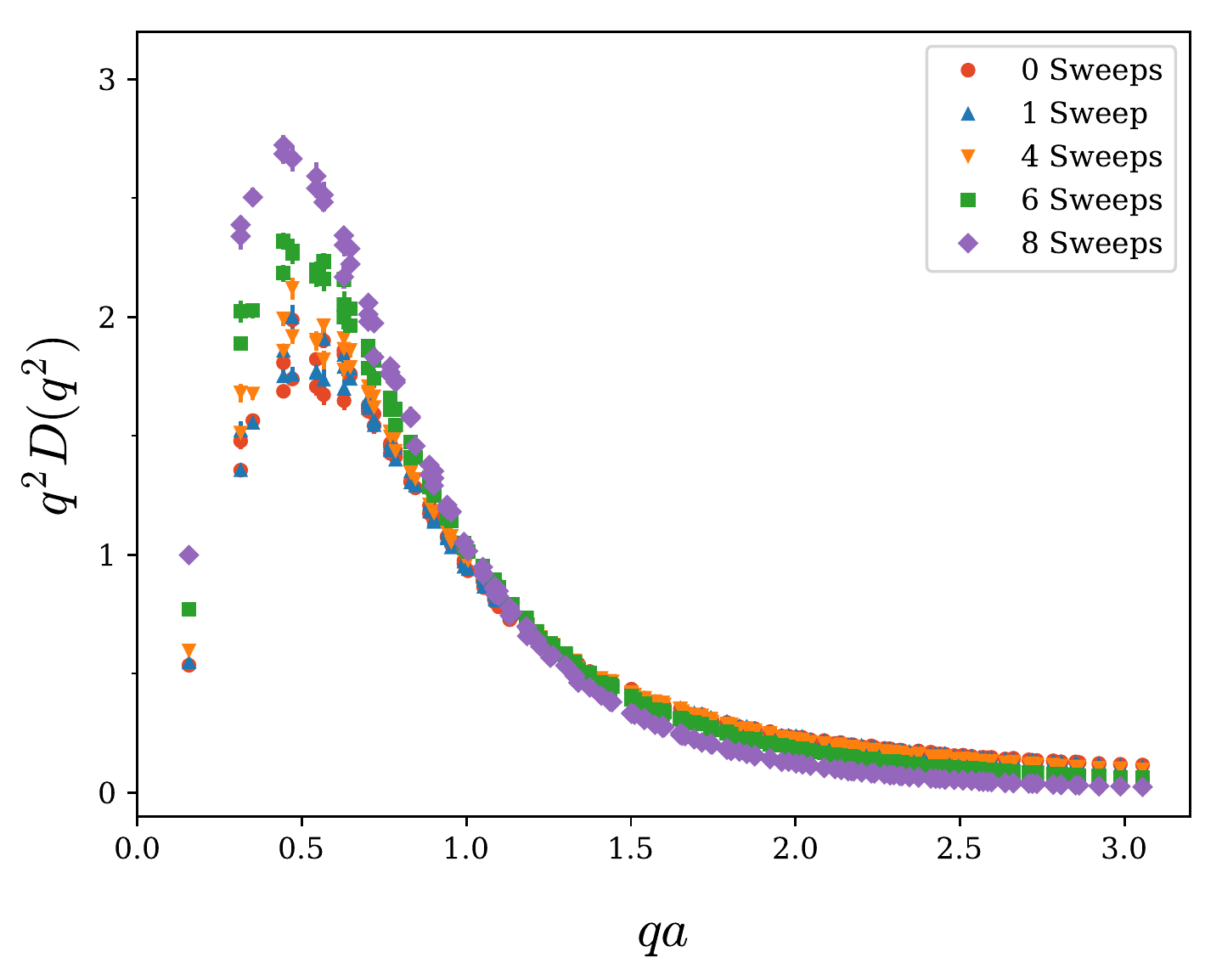}
\caption{\label{fig:1to10VO}The vortex only propagator after different sweeps of cooling. A trend similar to Fig.~\ref{1to10sweepscooling} is observed, with enhancement in the infrared and suppression in the UV region.}
\end{figure}

We perform the same analysis of the vortex only propagator under cooling as performed in section~\ref{sec:Smoothing} on the untouched propagator. Once again in gauge fixing, each sweep is preconditioned by the Landau gauge transformation of the previous sweep in descending order. The result of this analysis is shown in Fig.~\ref{fig:1to10VO}. This figure shows a similar change in the vortex only propagator when compared to the untouched propagator in Fig.~\ref{1to10sweepscooling}, with an enhancement in the infrared and suppression in the UV modes. The UV suppression is less noticeable in this case due to the prior removal of short range effects brought about by the vortex identification.

%
%
We observe that the vortex only and untouched propagators in Fig.~\ref{10sweepscooling} resemble the gluon propagator under a differing number of sweeps of cooling, as shown in Fig.~\ref{1to10sweepscooling} and Fig.~\ref{fig:1to10VO}. The vortex only propagator has a peak that sits below the untouched propagator, and the untouched propagator is further suppressed in the $qa\approx 1.5$ region. Following the trend in Fig.~\ref{1to10sweepscooling} and Fig.~\ref{fig:1to10VO}, this indicates that further cooling on the vortex only propagator would align it with the untouched propagator. This follows from an understanding that the vortex-only configurations are initially much rougher than their untouched counterparts~\cite{Trewartha:2015nna}, and should therefore require additional cooling to obtain agreement with the untouched configurations.\\
\begin{figure}[tb]
\centering
\includegraphics[width=\linewidth]{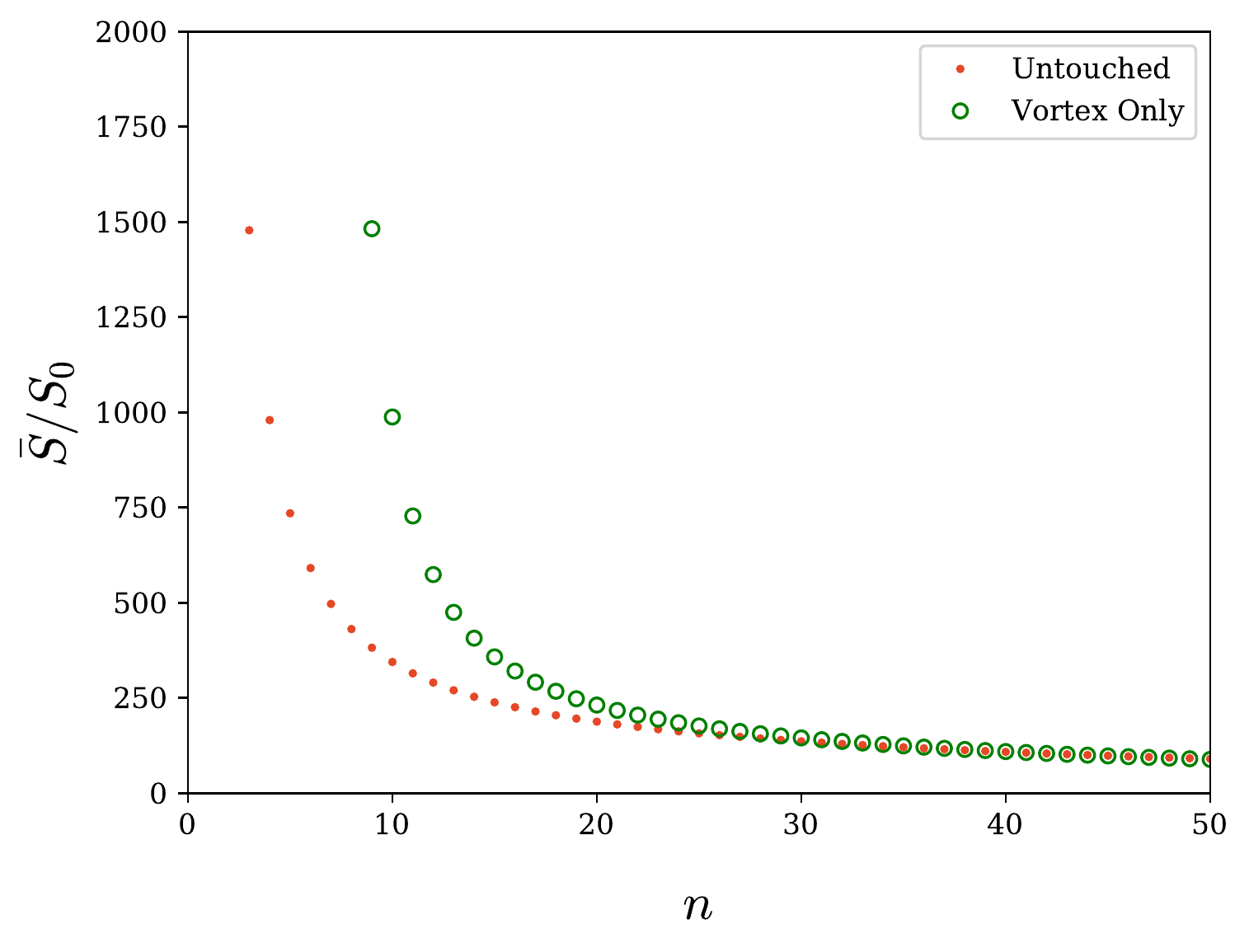}
\caption{\label{ActionMatch}The average action calculated on the untouched and vortex-only configurations as a function of cooling sweeps, $n$. The vortex only configurations are initially rougher than the untouched, as evidenced by the higher average action.}
\end{figure}
\begin{table}[tb]
\caption{\label{actionmatch}Comparison of the number of cooling sweeps on the untouched ($n_U$) and vortex only ($n_{VO}$) configurations required to match the average action.}
\begin{ruledtabular}
\begin{tabular}{cccc}
$n_U$ & $\bar{S}/S_0$ & $n_{VO}$ & $\bar{S}/ S_0$\\
\hline\\
5 & $734.83$ & 11 & $727.67$\\
10 & $344.22$ & 15 & $357.68$\\
15 & $238.21$ & 20 & $231.19$\\
20 & $187.55$ & 24 & $184.68$\\
25 & $156.92$ & 28 & $155.72$\\
30 & $135.91$ & 32 & $135.61$\\
35 & $120.29$ & 36 & $120.66$\\
40 & $107.08$ & 40 & $109.02$\\
\end{tabular}
\end{ruledtabular}
\end{table}
\begin{figure}[tb]
\centering
\includegraphics[width=\linewidth]{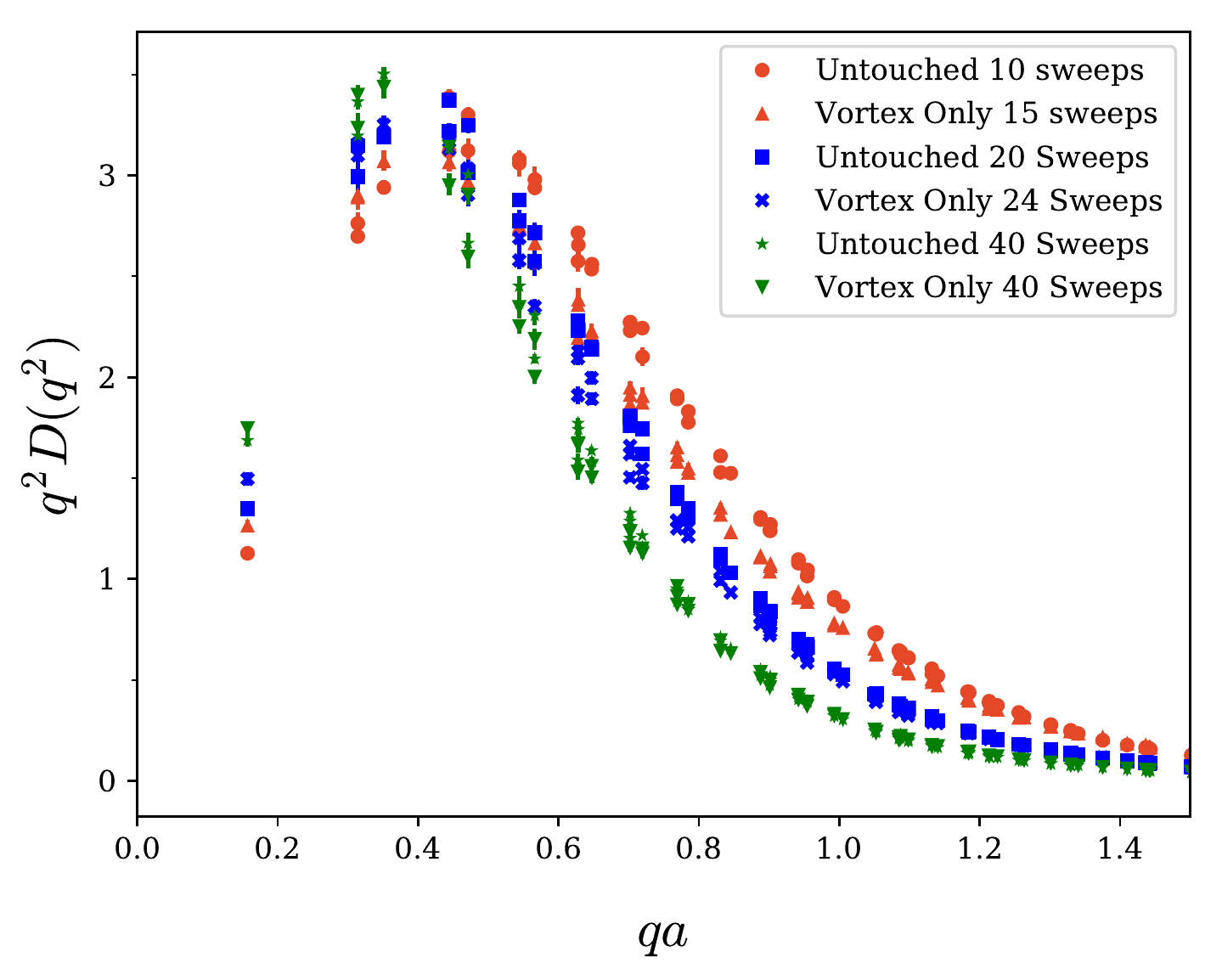}
\caption{\label{UVOActionMatch}Comparison of the gluon propagator on the untouched and vortex only configurations after tuning the number of cooling sweeps to best match the average plaquette action. This procedure gives a much better agreement in the shape of the gluon propagator from the two configurations.}
\end{figure}
We take the average $\mathcal{O}(a^4)$ three-loop improved action of the lattice divided by the single instanton action $S_0=\frac{8\pi^2}{g^2}$, denoted $\bar{S}/S_0$, to be a measure of roughness. We observe that for $n<20$ cooling sweeps the vortex-only configurations have a significantly higher action than their untouched counterparts after the same number of sweeps of cooling, as illustrated in Fig.~\ref{ActionMatch}. We therefore seek to find the number of sweeps required to best match the action between the vortex-only and untouched configurations. The results of this procedure are shown in Table \ref{actionmatch}. If we now plot these matched configurations, we obtain the results shown in Fig.\,\ref{UVOActionMatch}. Here we have truncated the plot at large $qa$ to better show the agreement in the mid-$qa$ region. By matching the actions as closely as possible with an integer number of cooling sweeps, we see that there is a better agreement between the untouched and vortex-only gluon propagators.

There is now a significant body of evidence that centre vortices contain the essential degrees of freedom of the Yang-Mills vacuum, such that the application of smoothing enables the recreation of the major features of QCD~\cite{Bertle:2001xd,Trewartha:2015ida,Trewartha:2015nna,Trewartha:2017ive,DelDebbio:1998luz}. Noting that sufficient smoothing of a vortex-only field generates a topological background of instanton-like objects, we can regard the thin centre vortices as the seeds of instantons. The smoothing process that is applied on the vortex-only configurations raises a question regarding the precise role of vortices in the restoration of the infrared propagator; is it simply the presence of (sufficiently smoothed) vortices or is it more indirectly the reformation of the instanton background? If we examine Fig.~\ref{10sweepscooling}, we see that after the application of 10 sweeps of cooling the vortex-only propagator has the appropriate qualitative infrared behaviour. Comparing with previous work, in particular Fig.~7 within Ref.~\cite{Trewartha:2015ida} which shows the typical distribution of the instanton radius against the topological charge at the centre, we can see that after only 10 sweeps of cooling the vortex-only distribution still deviates significantly from the ideal theoretical instanton relationship. This suggests that it is the smoothed centre vortices that are directly responsible for the infrared structure of the gluon propagator.


\section{Conclusions}
 
Through investigation of the Landau gauge gluon propagator on original, vortex removed and vortex only backgrounds, we have shown that centre vortices play a significant role in the structure of the gluon propagator. Vortex identification partitions the gluon propagator into short range strength on the vortex removed configurations and long range strength on the vortex only configurations. We also demonstrated that this partitioning is consistent with the vortex modified gauge potential representing orthogonal components of the original potential. Although the vortex only propagator does not exhibit the full infrared strength of the original configurations, it is clear that centre vortices encode much of the long range physics.

We then investigated the effect of smoothing on the gluon propagator, and determined that both cooling and over-improved smearing produce similar suppression of high frequency modes and amplification of infrared behaviour. After applying smoothing, the untouched and vortex-only configurations are brought closer together, and by using the average action as a measure of roughness we see that it is possible to recover the strength of the propagator on the vortex-only configurations when compared to the untouched configurations. The accuracy to which the smoothed vortex-only configurations are able to recreate the gluon propagator on similarly smooth original configurations is remarkable, as illustrated in Fig.~\ref{UVOActionMatch}.

This work motivates further exploration of the impact of centre vortices in full QCD, where we would anticipate infrared screening of the propagator~\cite{Bowman:2004jm}. It would also be of interest to investigate whether improved vortex identification can account for the discrepancy between the untouched and vortex only propagators. Additionally, there has been work to develop smoothing techniques that explicitly grow thin $SU(2)$ vortices by spreading the centre flux across larger Wilson loops~\cite{Hollwieser:2015koa}. The consequences of this novel smearing method in $SU(3)$ would be an interesting area for future work.

As discussed in Ref.~\cite{Trewartha:2015ida}, it is currently unknown whether the number of smoothing sweeps required to reproduce QCD properties increases or remains constant as the lattice spacing approaches the continuum limit. The former case implies that the purpose of smoothing is to grow thin vortices into thick vortices that have a similar size to those found on the original configurations; whereas the latter case indicates that smoothing is merely required to remove the roughness of the vortex-only configurations. As a result, we wish to investigate the impact of centre vortices and smoothing in the continuum limit. The results of this work contribute further numerical evidence that centre vortices are the fundamental mechanism underpinning QCD vacuum structure.

\acknowledgments

This work was supported with supercomputing resources provided by the Phoenix HPC service at the University of Adelaide. Additional computing resources used to assist this investigation were provided by the National Computational Infrastructure (NCI), which is supported by the Australian Government. This research was supported by the Australian Research Council through Grants No.\ DP140103067, DP150103164, and LE160100051.

\appendix*

\section{Notes on the gluon propagator}

Here we briefly recall the derivation of the expression used to calculate the momentum space gluon propagator on the lattice~\cite{Zwanziger:1991gz,Cucchieri:1999sz,Langfeld:2001cz}. It is instructive to reference the coordinate space form, 
\begin{equation}
D^{ab}_{\mu\nu}(x) = \langle A^a_\mu(x) \, A^b_\nu(0)\rangle.
\end{equation}
The propagator in momentum space is simply related by the discrete Fourier transform,
\begin{equation}
D^{ab}_{\mu\nu}(p) = \sum_x e^{-ip\cdot x} \langle A^a_\mu(x) \, A^b_\nu(0) \rangle. 
\end{equation}
Noting that the coordinate space propagator $D^{ab}_{\mu\nu}(x-y)$ only depends on the difference $x-y,$ we can make use of translational invariance to average over the four-dimensional volume to obtain the form for the momentum space propagator given in Eq.~\ref{eq:gluonProp},
\begin{align}
D^{ab}_{\mu\nu}(p) &= \frac{1}{V}\sum_{x,y} e^{-ip\cdot x}\langle A^a_\mu(x+y) \, A^b_\nu(y) \rangle \nonumber \\
                &= \frac{1}{V}\sum_{x,y} \langle e^{-ip\cdot (x+y)} A^a_\mu(x+y) \, e^{+ip\cdot y}A^b_\nu(y) \rangle \nonumber \\
                &= \frac{1}{V}\langle A^a_\mu(p) \, A^b_\nu(-p) \rangle. \label{eq:gluPropxtop}
\end{align}
On a discrete lattice we prefer to make use of the mid-point
definition of the gauge potential, as this yields a symmetric local Landau gauge condition~\cite{Alles:1996ka},
\begin{equation}
\Delta(x) = \sum_\mu A_\mu(x+\hat{\mu}/2) - A_\mu(x-\hat{\mu}/2) = 0.
\end{equation}
This local gauge condition in momentum space is then
\begin{equation}
\sum_\mu 2i\sin\frac{p_\mu}{2} A_\mu(p) = 0,
\end{equation}
which is free from $\mathcal{O}(a)$ errors. The Landau gauge fixing functional that we use to transform the links $U_\mu(x)$ is further improved by taking a combination of one- and two-link terms to eliminate $\mathcal{O}(a^2)$ errors~\cite{Bonnet:1999mj}.

The scalar propagator is obtained from the Lorentz diagonal components of the gluon propagator. In the case $\mu = \nu,$ it is straightforward to replace $y \to y + \hat{\mu}/2$ in equation~(\ref{eq:gluPropxtop}) to derive  
$D^{aa}_{\mu\mu}(p)$ in terms of the mid-point definition of the $A_\mu$ fields. The propagator itself is calculated directly from the potential in momentum space to avoid the problem of statistical noise in the coordinate space propagator at large separations $|x-y|$~\cite{Zwanziger:1991gz,Cucchieri:1999sz,Langfeld:2001cz}.

\bibliographystyle{apsrev4-1}

%

\end{document}